# Atomic Level Strain Induced by Static and Dynamic Oxygen Vacancies on Reducible Oxide Surfaces

Piyush Haluai, Tara M. Boland, Ethan L. Lawrence, Peter A. Crozier*

*School for Engineering of Matter, Transport & Energy, Arizona State University, 501 E. Tyler Mall, Tempe, Arizona 85287 (United States)*

*Corresponding author: crozier@asu.edu

**Abstract:**

Surface strain often controls properties of materials including charge transport and chemical reactivity. Localized surface strain is measured with atomic resolution on (111) ceria nanoparticle surfaces using environmental transmission electron microscopy under different redox conditions. Density functional theory (DFT) coupled with TEM image simulations have been used to interpret the experimental data. Oxygen vacancy creation/annihilation processes introduce strain at the surface and near surface regions of the cation sublattice. Both static and fluxional strain maps are generated from high resolution images recorded under varying reducing conditions. While fluxional strain is highest at locations associated with unstable vacancy sites, highly inhomogeneous static strain fields comprising of alternating tensile/compressing strain is seen at the surface and subsurfaces linked to the presence of stable oxygen vacancies. Interestingly, both stable and unstable oxygen vacancies are found within a few atomic spacings of each other on the same surface. The static strain pattern depends on the ambient environment inside the TEM.

# Introduction:

Strain engineering has emerged as a promising field for modifying/altering the structure-property relationships in materials at the atomic level [1]. Interfaces and surfaces play a crucial role in determining many materials functionalities and characterizing the strain or structural distortions at high spatial and temporal resolutions is essential for developing a fundamental understanding of how strain effects the properties of the material [2] [3]. For example, surface strain can regulate and control surface diffusion processes and can change the chemical reactivity of a surface, such as, enhancement of oxygen reduction activity [4] [5], catalytic properties (light-off temperature and attainable activity) [6]–[8], adsorption energy at the surface [9], strain-induced corrosion [10] etc. In reducible oxides, point defects such as oxygen vacancies distort/strain the cation sub-lattice and influence surface properties such as reactivity, and structural stability. The concentration of oxygen vacancies also alters the structure of these materials. For example, in ceria, which is a high-symmetry fcc fluorite structure, changes to a low-symmetry bcc structure beyond a certain vacancy concentration [11]. Since the concentration of oxygen vacancies can be varied with different materials processing, characterizing the atomic-level variation in strain under different conditions with transmission electron microscopy will provide fundamental new insights into the dynamic behavior of the associated surface strain fields.

Here we investigate the surface strain and associated oxygen vacancy behavior on ceria ($CeO_2$) nanoparticle surfaces. Ceria is an exemplary reducible oxide with technological importance. Ce cations change their oxidation state with the creation/annihilation of oxygen vacancies and the material has a good structural stability [12]. Due to its ability to exchange lattice oxygen with its surrounding environment, ceria and its doped counterparts have applications in various fields such as catalysis and electrodes for the solid oxide devices, biomedicine etc. [13].

The degree of strain tuning can be varied by changing particle size and shape, non-stoichiometry, ambient environment [14] etc. The presence of surface oxygen vacancies in ceria has been investigated by many researchers [14]–[17]. The atomic level dynamics of vacancy creation/annihilation associated with oxygen exchange is also important in surface redox processes such as catalysis, and has been investigated with *in situ* electron microscopy [18] [19].

In reducible oxide systems, it is helpful to think about strain and structural stability in terms of the associated anion and cation sublattices. In general, the anion sublattice will be more dynamic or fluxional due to the creation and annihilation of oxygen vacancies associated with transport and surface exchange. The fluxionality of the anion sublattice can be described in terms of the variation in the occupancy and location of each anion site. One consequence of the ionic bonding in these ceramic oxides is that changes in the anion occupancy will influence the positions of the neighboring cations. As we will show, much of the cation sublattice strain in our observations is associated with anion activity. Moreover, if the coordination environment around the cation is low, the local structure may become unstable resulting in cation transport. This may happen in the presence of locally high concentrations of oxygen vacancies at cation sites such as step edges or adatoms. Understanding the local relationship between fluxionality, vacancy creation/annihilation and structural stability/instability is important for many applications and can impact device aging and durability. Note that in this manuscript, we are not talking about thermal vibrations, which occur on very fast timescales.

Strain measurements provide a convenient descriptor of relative lattice distortions for both bulk and surface phases. Traditional transmission electron microscopy (TEM)-based strain measurement techniques have been employed to measure the average distance between neighboring atomic columns in an image, providing a description of the defects at the surface. In CeO$_2$, the Ce sublattice columns can be located with higher precision than the anion sublattice because of the stronger TEM imaging signal in comparison with the much lighter and dynamic oxygen.

Our previous work on CeO$_2$ at higher time resolution showed that at some surface sites, cations can shift by around $10 – 20$ pm between frames with a displacement frequency, at room temperature, of $5 – 20$ Hz. This is a chemical effect resulting from the relaxation of local cations during the creation and annihilation of oxygen vacancies at nearby anion sites [18]. This fluxional behavior results in a diffuse appearance of active cations in images recorded with longer exposure times. The resulting time-averaged mean-square displacements in the cation sublattice associated with this phenomenon can be described using a concept called *fluxional strain*. We have used the concept of fluxional strain to provide a structural descriptor of the surface catalytic activity for CO oxidation [19]. The fluxional strain is highest for cations near the most active oxygen vacancy sites. To avoid confusion, we refer to the traditional definition of strain as static strain to emphasize the fact that during a typical observation time (on the order of seconds), the static strain field will vary only slowly with time.

Our analysis explicitly shows that the surface oxygen vacancies can be classified into two categories: stable and unstable vacancies. Stable vacancies are associated with stable static cation strain fields and may lead to long-lived metastable surface structures. The fluxional strain identifies the location of unstable oxygen vacancy configurations leading to continuous changes in local surface structure during the observation period. These two types of surface lattice distortions co-exist simultaneously and within a few atomic spacing of each other. The primary scientific goal of this manuscript is to characterize and understand these complex surface distortions and how they change in the presence of different ambient environments.

In this work, we employed *in situ* transmission electron microscopy with varying ambient atmosphere and electron dose to induce changes in the oxygen vacancy distribution in a $CeO_2$ nanoparticle. In an electron microscope, the electron scattering from the heavier Ce cation columns is much stronger than the scattering from the lighter oxygen (O) anion columns. Consequently, higher precision measurements can be performed on the cation sublattice to map the strain on or near the nanoparticle surfaces. Many TEM techniques are available to measure strain using both reciprocal and real space techniques [6] [20] [21]; here we employ real space techniques. Strain maps were generated from the same area of the nanoparticle after changing the ambient environment (oxidizing/reducing condition at room temperature) inside the microscope. We also employed density functional theory (DFT) coupled with TEM image simulation to create models of surface vacancies in ceria to facilitate the interpretation of the strain maps.

## Materials and Methods:

$CeO_2$ nanoparticles were synthetized using a hydrothermal method [22]. The synthesized powder was then calcined at 350°C for 2 hours and transferred through air to the microscope for characterization. High resolution phase contrast TEM imaging was performed using a Thermo Fisher Titan aberration-corrected environmental transmission electron microscope (AC-ETEM) operating at 300 kV. Negative $C_s$ imaging was employed to enhance the contrast from both the anions and cations [23]. The ETEM capabilities of the Titan microscope allow gas to flow inside the microscope enabling imaging of structural changes happening in different atmospheres. Images were recorded at room temperature using a direct electron detection camera (Gatan K2) operated in counting mode. To study the effect of different redox conditions on surface strain, the same nanoparticle was exposed, *in situ*, to 3 different conditions respectively:

***Condition A:*** Imaging in vacuum (initial condition). A careful and systematic study revealed that dose rates of 5 x $10^3$ eÅ$^{-2}$s$^{-1}$ (electrons per Ångstroms squared per second) or less minimizes electron beam induced radiation damage and thus was chosen for imaging at the neutral redox condition [18].

***Condition B:*** Imaging in 0.5 Torr of $O_2$ (oxidizing). The dose rate for the data acquisition was kept the same that of the previous condition of 5 x $10^3$ eÅ$^{-2}$s$^{-1}$.

***Condition C:*** Imaging with higher electron dose (reducing). The dose was increased by a factor of ~1000 to 5 x $10^6$ eÅ$^{-2}$s$^{-1}$. Such a high dose results in partial reduction of $CeO_2$ and the introduction of a higher concentration of oxygen vacancies [24]–[26].

The three conditions above were all at room temperature to slow bulk oxygen diffusion kinetics and provide a relatively stable 3D structure during the image acquisition time of 1s, enabling strain to be determined with reasonable precision. Here the primary focus is on the dominant (111) surfaces of the $CeO_2$ nanoparticle. A key aspect of our approach is determining the surface strain field from the same nanoparticle under the 3 different *in situ* conditions. This allows a detailed comparison of the changes in strain around the same defects (such as surface steps) with differing conditions. *In situ* comparison also provides information on the specifics of cation surface diffusion resulting from the introduction of high concentrations of oxygen vacancies. Some changes in imaging conditions are inevitable (mostly due to nanoparticle tilt which is discussed later) when the redox condition is changed. However, the errors are manageable and deeper scientific insights into the structural transformations taking place can be derived from strain mapping at identical locations in 3 different redox conditions.

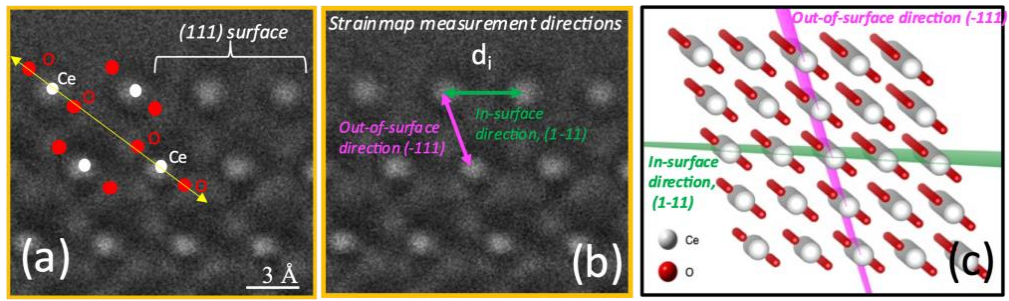

**Figure 1:** (a) High resolution phase contrast image of a (111) surface of the nanoparticle viewed along [110] zone axis showing Ce and O atomic column position and the "-O-Ce-O-O-Ce-O-" motif (marked by arrow) along the body diagonal ([100] direction) of the parallelogram formed by the cation sublattice. (b) in-surface and out-of-surface directions are shown on the nanoparticle surface along which the strain measurements were carried out. $d_i$ refers to distance measured between two neighboring columns. (c) a model of $CeO_2$ at [110] zone axis orientation. The structure is tilted for ease of visualization of the atomic planes labeled. (111) planes along which the strain is measured in the experimental data is shown and matched with the color of (b).

**Figure 1a** shows a high-resolution 1 s exposure image of (111) surface of the nanoparticle viewed along [110] zone axis showing Ce atomic columns as intense white dots and the oxygen (O) columns as weaker white dots in a "-O-Ce-O-O-Ce-O-" type atomic arrangement visible along the body diagonal (marked by arrow) of the parallelogram motif formed by the cation sublattice at the nanoparticle surface/subsurfaces ([100] direction). Due to the higher signal-to-noise ratio (SNR), static surface strain measurements (relative to the nanoparticle bulk) are performed on cation (Ce) columns by comparing the spacings between neighboring columns ($d_i$) at the surface or near surface region to the symmetrically equivalent bulk spacing, ($d_o$), measured from a 10 x 5 block of unit cells at least 3-4 layers away from the surface columns for each condition (A, B, and C). There are oxygen vacancies and associated static strains on the electron beam entrance and exit surface of the sample regions where the value of $d_0$ is determined. However, these tensile and compressive strains will be averaged out when $d_o$ is determined over a large number of unit cells. The strain is measured along (111) and (-111) planes parallel to and at ~70º to the surface direction, respectively (see **figure 1b**). Along the [110] zone axis orientation, the equivalent (111) planes are the corresponding in-surface and out-of-surface directions as marked by the arrows in **figure 1b**. A model of CeO$_2$ (**figure 1c**) shows the planes along which the strain fields were measured.

To determine the coordinates of the atomic columns to sub-pixel accuracy, 2D elliptical Gaussians were fitted to each cation (Ce) column [27] allowing the distance between adjacent cation columns to be determined. The cation column distances were measured at different locations on the nanoparticle (surfaces and subsurfaces). The % static strain was then calculated as:

$$\% \ static \ strain = \ 100 * \left(\frac{d_i - d_0}{d_0}\right) \quad --- \textbf{equation (1)}$$

Although the static strain can show the distortions in the lattice from a 1 sec image exposure, it is an incomplete description of the local distortions since it does not capture cation column fluxionality. To understand why, it is useful to consider a simple example. Suppose we have two locations on the surface with no static strain but different fluxional behavior. At one location and time t/2, the cation is at displacement of +δx away from its mean position, and a time t/2, it is at a displacement -δx from its mean position. At the other location and time t/2, the cation is at displacement +δy (δy ≠ δx) away from its mean position and a time t/2 at a displacement -δy from its mean position. Over a measurement time t, the mean displacement of the two cations from their relaxed positions is the same i.e., the static strain is the zero, but the fluxionality is different (see **fluxional strain formulation**, **figure S1 in supplemental**).

To address fluxionality, we use the concept of fluxional strain [19] defined as:

$$\% \ fluxional \ strain = \ 100 * \left(\frac{\sigma_i - \sigma_o}{d_0}\right) \quad --- \textbf{equation (2)}$$

where $\sigma_i$ is the standard deviations of the Gaussians fitted on each atomic column positions, $\sigma_0$ is the average standard deviation of the Gaussian fitted on the columns at the bulk, and $d_0$ is the symmetrically equivalent bulk spacing between atomic columns measured in each condition (A, B, and C).

DFT based simulations of oxygen vacancy induced ceria structures were also investigated. All density-functional theory calculations used the projector-augmented wave method [28], [29] as implemented in the plane-wave code VASP ([30], [31], [32], [33]). All calculations included spin-polarization and the generalized gradient approximation (GGA) with the Perdew-Burke-Ernzerhof (PBE) [34] exchange-correlation functional. In addition, the strong correlation effects of the Ce 4f electrons were treated within GGA using the Hubbard U correction (GGA+U) formulated by Dudarev et al [35]. An on-site Coulomb interaction, $U_{eff}$ = 5 eV, was used for Ce, as determined by Dholabhai et al. [36] as well as many others ([37], [38], [39], [40]), to provide a better fit with the experimental band gap ($E_{\text{gap}}$), lattice parameter ($a_0$), and bulk modulus ($B_0$) compared to traditional GGA methods. A plane wave cutoff energy of 520 eV was used for all cases using the block Davidson [41] minimization algorithm with an energy cutoff of $10^{-6}$. This cutoff energy was sufficient to converge the forces [42] acting on each ion to 0.02 eV/Å per atom or better. For a 2x2x2 supercell of bulk ceria, we find that $E_{\text{gap}}$= 2.0 eV, $a_0$=5.494 Å, and $B_0$=180.59 GPa and the corresponding experimentally measured values are $E_{\text{gap},(O2p)\rightarrow Ce\,(4f)} =$ 3.0 eV [43], $a_0$=5.411 Å [44], and $B_0$=204-236 GPa. ([45], Gerward and Olsen 1993). The chosen value of $U_{\text{eff}}$ correctly describes the localization of the $4f$ electrons on the nearby Ce atoms–unlike traditional GGA which results in delocalized electrons on all cerium ions in the lattice. Additional information regarding the creation of the surface slabs and relevant simulation parameters are discussed in the SI (see description of **figure S4 in supplemental**).

TEM images were simulated using the Dr. Probe software [46] from the DFT relaxed models (with simulation cell parameters as a= 13.25 Å, b= 38.06 Å and c=7.7 Å (along the electron beam direction), with an average of 3 oxygen columns and 3 cerium columns along the electron beam direction respectively) using the same imaging parameters (pixel size= 8 pm, defocus= 3 nm, and spherical aberration coefficient= -13 μm) mentioned above. Strain maps were generated following the same procedure mentioned above.

## Results and Discussion:

### Experimental observations

#### *Condition A: Strain maps in vacuum (initial condition)*

**Figure 2a** shows a stepped (111) surfaces from a $CeO_2$ nanoparticle (100 nm X 50 nm in size) and although the cation (Ce) columns appear to be visible with very good signal-to-noise, the oxygen (O) columns suffer from poor signal-to-noise due to their lower atomic number (Z=8 for O as compared to 54 for Ce). Contrast changes arising from experimental parameters such as tilt and thickness variations (e.g., from surface to bulk) alter the appearance of the image from a simple projected potential map and complicate image interpretation. Such effects are particularly apparent in the lower left-hand region of **figure 2a**. However, since the primary interest of this study is the surface and local subsurfaces sites, thickness changes do not significantly impact determination of the surface strain. Detailed strain analysis is illustrated on this one nanoparticle but similar effects occur on many nanoparticle with varying size [47].

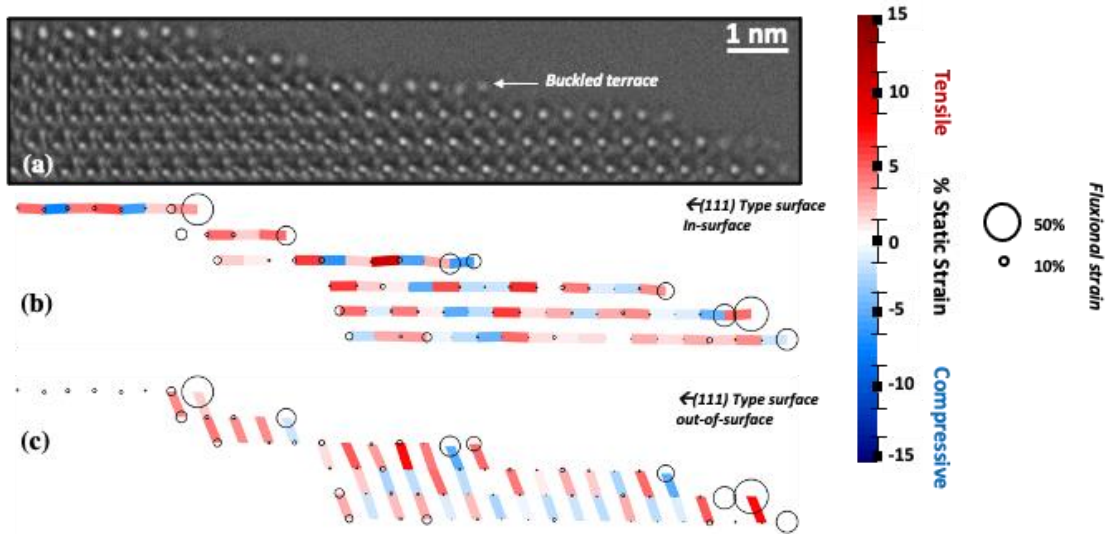

**Figure 2:** (a) *Condition A* image of ceria nanoparticle at [110] zone axis is displayed. Atomic columns (Ce, O) appear as bright blobs due to negative Cs imaging. (b) and (c) show in-surface and out-of-surface strain maps of the image in (a) is shown for (111) surfaces. Alternating tensile-compressive static strain fields are visible at the surface and subsurfaces of the nanoparticle in both cases. The buckled terrace shows the highest value of static strain. Fluxional strain is represented in (b) and (c) for each column by the black circles. It is maximum at the step edges and surfaces. The correlation between the circles and the fluxional strain value is shown in the right side. Color bar at the right of the image shows correlation between color and % static strain. Error bars in strain measurement are also indicated.

Some of the atomic columns visible at the surface appear highly diffuse due to fluxionality associated with highly active oxygen exchange sites discussed previously [18]. The fluxional strains are represented in **figure 2b** and **2c** by means of variable sized black circles at each atomic column, reaching maximum values mostly at the step edges (roughly between 15-30%). The surfaces and subsurfaces typically show fluxional strain values of a few percent.

The static in-surface and out-of-surface strains, from the average position of the cations, are also shown in **figure 2b & 2c**. The strain maps show a highly inhomogeneous strain-field with a characteristic alternating tensile-compressive strain between adjacent column pairs and are quite different between different terraces. The in-surface strain can be as high as 6% compressive to 10% tensile at different parts of the surfaces. The out-of-surface strain map is similar, showing both tensile and compressive strains in the range ~2-8%. The highest degree of strain (both in-surface and out-of-surface) is associated with the region of the (111) surface that is buckled (labeled in **figure 2b)**. At room temperature, buckling of ceria films with oxygen vacancies are spontaneous and have been observed [48]. Interestingly, there is no strong correlation among different sites between fluxional strain and static strain.

***Condition B: Strain maps in oxygen environment (oxidizing condition)***

**Figure 3** displays the image and corresponding strain maps from the same nanoparticle in the presence of 0.5 Torr of $O_2$. Nanoparticles may tilt slightly on introduction of gas to the reaction cell, and this is clearly visible at the left part of the image showing elliptical cation columns compared to the circular shape in the image from *Condition A* (**figure 2**). Many of the surface Ce columns also show a pronounced diffuse contrast indicating a high degree of fluxionality. However, we are still able to measure mean variations in column spacings and estimate strain in the thinner surface regions.

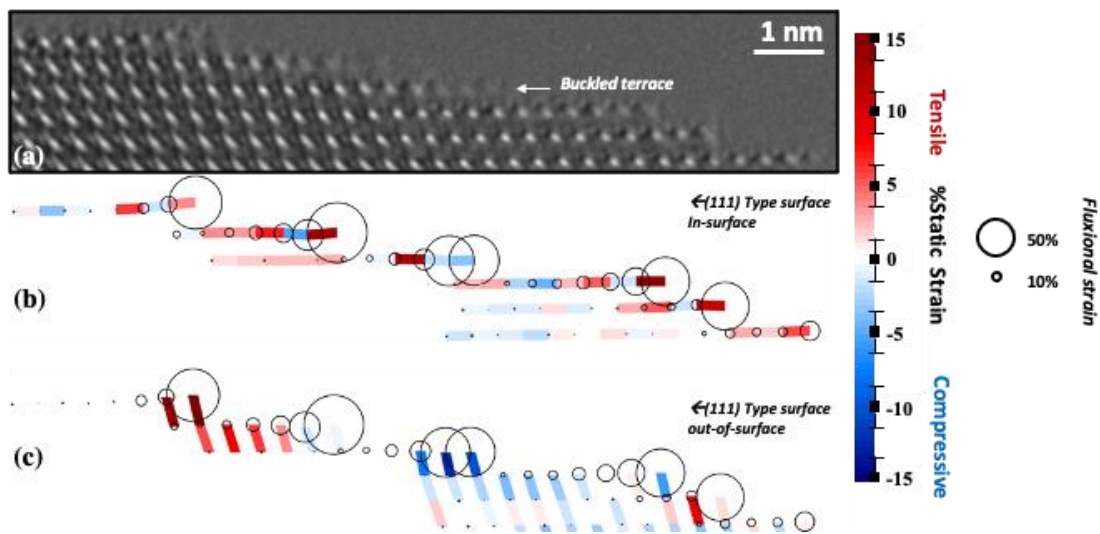

**Figure 3:** (a) Image of the ceria nanoparticle at [110] zone axis in 0.5 Torr of $O_2$ (*Condition B*). Effect of nanoparticle tilt due to the presence of gas inside the microscope is visible in the shape of cation columns (from circular to elliptical). (b) in-surface strain map of the image in (a) is shown for (111) surfaces. Alternating tensile-compressive strain fields are visible mostly at the surface of the nanoparticle. The strain fields are less intense towards the bulk of the nanoparticle as compared to figure 2b. The buckled terrace still shows high degree of tensile-compressive strain. (c) shows out-of-surface strain fields at the surfaces with different surfaces showing tensile and compressive strain. The strain is maximum at the surface and drops as we move inside the nanoparticle. Fluxional strain is represented in (b) and (c) for each column by the black circles. It is maximum at the step edges and surfaces and is higher than the vacuum condition. The correlation between the circles and the fluxional strain value is shown in the right side. Color bar at the right of the image shows correlation between color and %strain. Error bars in strain measurement are also indicated.

In this case, fluxional strain values are significantly higher (~10-50%) and now extend from the step edges across the terraces with decreasing value. The static strains for the nanoparticle in the $O_2$ atmosphere range from ~1-13% with the details of the strain field being different from the initial state (*Condition A*). The strain is mostly concentrated at the surface and decays more rapidly into the bulk. An alternating tensile and compressive static strain pattern is still present at some sites. Strain fields vary along different (111) terraces with the highest strain being present at the previously buckled terrace. Three out of six step edges in **figure 3** show tensile strains above ~5%. Interestingly, when the terrace is strained (due to buckling) the static strain at the step edge is much smaller, but the fluxional strain is still high.

### *Condition C: Strain maps in vacuum at high electron beam dose (reducing condition)*

The image (**figure 4a**) shows a better signal-to-noise compared to **figure 2a** and **3a** due to the increased electron dose. Some additional tilt has occurred as evidenced by elliptical shape of the cation columns at the surface and the bulk as well. The cation columns associated with the buckled terrace in the lower dose image (**figure 2b and 3b**) are missing in the high-dose image (the columns were present in the image from *Condition B*) indicating that there's been cation transport at these surfaces.

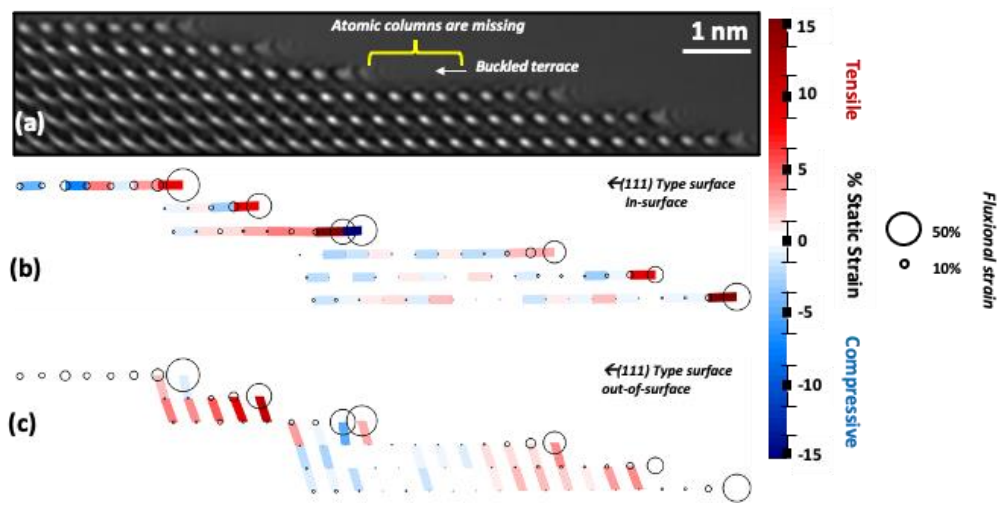

**Figure 4:** (a) High-dose image of the ceria nanoparticle at [110] zone axis (*Condition C*). Better image contrast (as compared to the low-dose images) is clearly visible. The Ce atomic columns are showing much pronounced tilt effects compared to previous images. (b) in-surface strain maps of (111) surfaces showing alternating tensile-compressive strain fields at the surface. The region labeled "buckled terrace" in the low-dose images (figure 2 and 3) shows very high degrees of tensile-compressive strain fields, but the terrace seems to be missing atoms beyond the buckled part. (c) out-of-surface strain fields are small for most of the surfaces apart from the buckled one. Fluxional strain is represented in (b) and (c) for each column by the black circles. It is maximum at the step edges and surfaces. The correlation between the circles and the fluxional strain value is shown in the right side. Color bar at the right of the image shows correlation between color and %strain. Error bars in strain measurement are also indicated.

Fluxional strain maps are similar to that of *Condition A* showing the presence of high degrees of strain at the step edge (value ~10-30%). Subsurfaces show fluxional strains typically of a few percent only.

Interestingly, the in-surface (111) static strain fields still show the alternating tensile-compressive pattern mostly at the surfaces. The previously buckled step edge shows the highest degree of strain (~17%) although most of the step edge sites still show high static strain values. Apart from the terrace with missing columns, most of the nanoparticle surfaces have small amount of out-of-surface strain (~1-4%).

## Discussion:

### *Random and Systematic Errors:*

Before discussing the scientific interpretation of the strain measurements, it is important to consider the errors in the measurements. The error for static strain measurement originates from two sources, random error associated with the noise fluctuations in the image and systematic error caused by changes in crystal tilt. The random error due to the presence of noise in the static strain measurement depends on the precision in the determination of the positions (X and Y coordinates) of the cation columns. To estimate the magnitude of this error, images of ceria were simulated, and Poisson noise added to match the SNR of the experiment. The simulations were performed with the same pixel size as the experimental images and 10 different noise realizations were generated (see **supplementary info, figure S2**). The atomic column positions were determined for each of the 10 noisy images using the same approach described in methods. The standard deviations, $\sigma_x$ and $\sigma_y$ were determined for the position of each cation column in the image. The total error ($\sigma_{X,Y}$) for each atomic column position was then determined as:

$$\Delta\sigma_{X,Y} = \sqrt{\sigma_x^2 + \sigma_y^2},$$

Thus, while measuring distance between any two cation columns the error is,

$$\Delta d_i = \sqrt{\Delta\sigma_{X,Y}^2 + \Delta\sigma_{X,Y}^2}.$$

In the experimental data, there are cation columns at the surfaces showing diffuse contrast compared to the bulk cations. So, it is necessary to estimate the error associated with sharp and diffuse atomic columns (shown in **supplementary info (figure S3)).** For a sharp cation column with 10 different noise realizations, the $\sigma_{X,Y}$ value is 0.9 pm. So, the error is, $\Delta d_i = \sqrt{0.9^2 + 0.9^2} =$ 1.2 pm (i.e., ~0.35% error in static strain measurement knowing the spacing in projection between two atomic columns along the (111) plane directions (331 pm)). While the same $\Delta d_i$ value for columns with diffuse appearance is 2.4 pm (i.e., ~0.7% error in static strain measurement). So, any single static strain measurement below 0.7% may be the result of random noise fluctuations in the image for surface columns (at the 1 standard deviation level, 1.4% with 2 standard deviation and 2% with 3 standard deviations (99% certainty)).

An estimate of the systematic error introduced by crystal tilt was also addressed with image simulation. Simulations were performed with added tilt on the 40 Å thick ceria model to get an approximate match to the experiment. A tilt of 1 degree along both of the (111) planes was applied, and the images were simulated (**figure 5**). The image shows smearing of the oxygen columns along with the cerium columns changing from a circular to elliptical shape. The strain maps derived from the simulated tilted structure shows an apparent average of ~1.5% alternating tensile/compressive strain fields for both in-surface and out-of-surface directions. Consequently, the error associated with crystal tilt can be estimated to be ~1.5%. The fluxional strain estimated from the simulations was <1% making systematic error negligible for the current analysis.

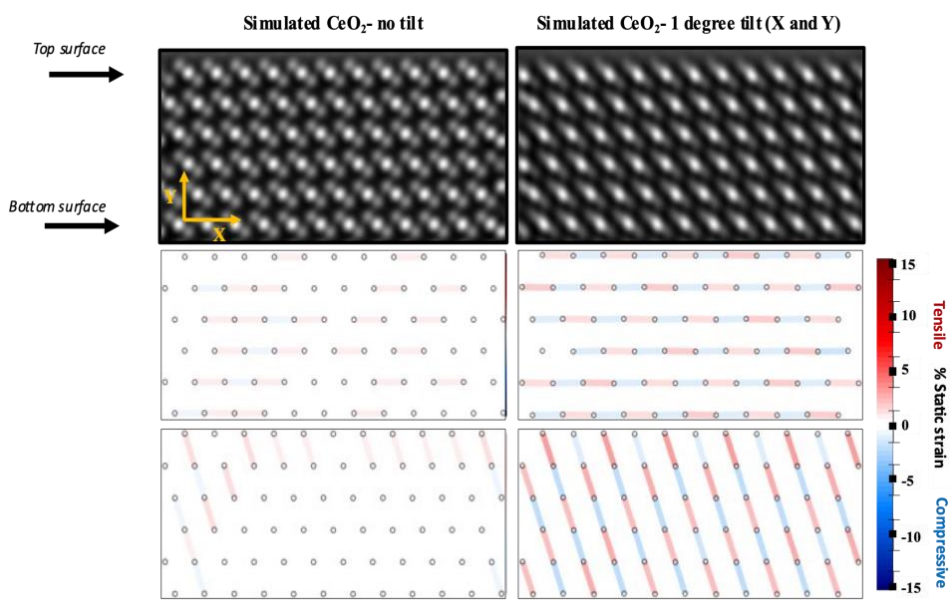

**Figure 5:** Simulated images of CeO₂ with added tilt. The simulated image shows streaking along the body diagonal of the cation sublattice. The strain maps also show alternating tensile and compressive strain fields. The degree of these strain values is on an average ~1.5%.

Random error in the fluxional strain measurement due to noise is carried out by measuring the standard deviation of each atomic column from the simulated images with added noise and then calculating the fluxional strain in each noise realization with respect to the noise free image using **equation 2**. The average fluxional strain with 10 different noise realizations is 1.3% with 1 standard deviation, 2.5% with 2 standard deviations, and 3.7% with 3 standard deviations (99% certainty).

***DFT-relaxed models with oxygen vacancies at the surface***

To further the interpretation of the static strain-fields in the experimental data, DFT simulations were carried out for structures with oxygen vacancies (single and double) at the surface

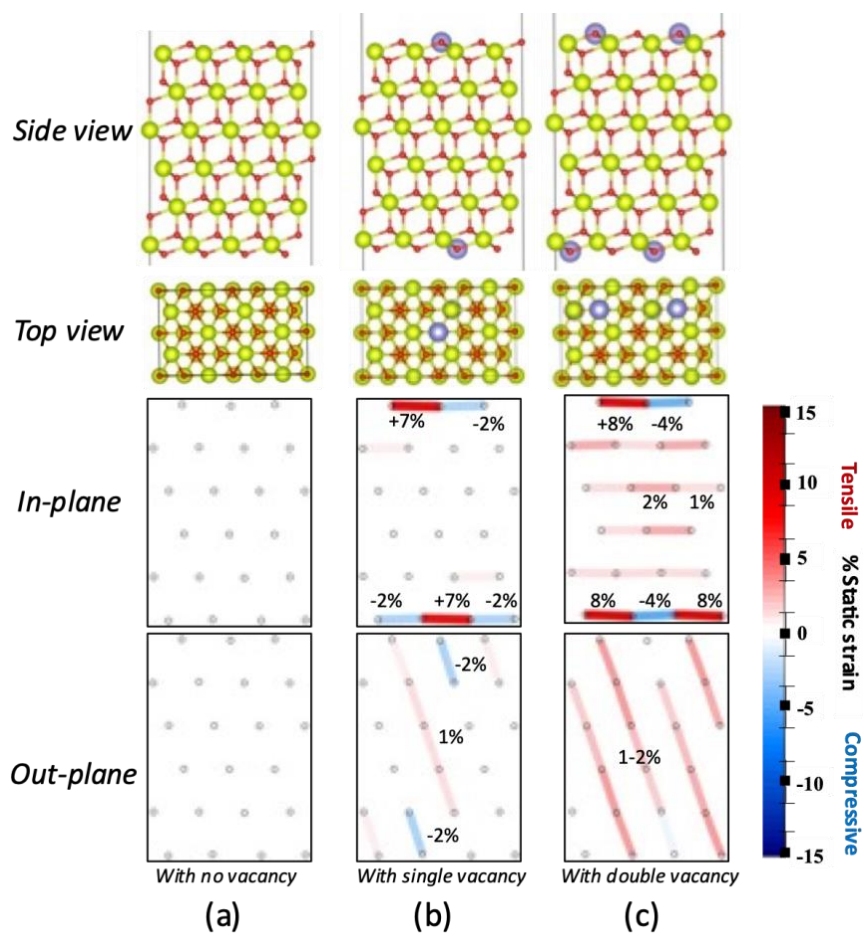

**Figure 6:** Simulated models of Ceria with (a) no vacancy, (b) single vacancy, and (c) double vacancy. The top two rows show the model structure with oxygen vacancy marked by the blue circle. The bottom two rows show the in-surface and out-of-surface strain maps for the models. In (a), there doesn't seem to be any strain fields present between the atomic columns. In (b) and (c) significant strain fields are present along both the directions. % Static strain values are also marked in the strain maps too. For the model in (c), the strain fields extend to the bulk layers as well. Fluxional strain isn't plotted in this image since the values are very small (<1%) and makes it difficult to locate the positions of the cation columns in the strain maps.

layer (top and bottom both) of the supercell structure (**figure 6**). After creating the oxygen vacancy,

the structure was relaxed, TEM images were simulated (**shown in the supplemental info, figure S4**), and corresponding static strain maps generated from the images using the method described before. Oxygen vacancies are marked by the blue circle in the figure. For the single vacancy structure, alternating tensile-compressive strain fields around the vacancy on the (111) surface (**figure 6**) was observed along in-surface direction (7% tensile and 2% compressive), while the out-of-surface direction mostly shows compressive strain (-2%) on the cation column which is connected to the oxygen vacancy. Upon increasing the number of vacancies at the surface to two (while maintaining stability of the model) as shown in the **figure 6,** the static strain fields increase but the characteristic tensile-compressive pattern remains. Also, ~1-2% static tensile strain is visible in the out-of-surface direction in the subsurfaces. Overall, increasing the number of vacancies in the surface (top and bottom) layers has increased the % static strain at the surface and subsurfaces. All the measurements are normalized using the bulk spacing from the structure with no vacancies.

Previous reports have suggested that creation of an oxygen vacancy in ceria causes the two nearest neighbor cation columns to move away from each other (thus increasing the distance between them i.e., tensile strain and reducing the distance between their next neighbors along the surface, i.e., compressive strain) [18]. Thus, the DFT simulations show that stable oxygen vacancies on the surface of $CeO_2$ lead to a characteristic tensile/compressive strain field for neighboring cations. This tensile/compressive strain motif is the fingerprint of stable surface oxygen vacancies.

### *Stable and Unstable Oxygen Vacancies on the Surface*

The DFT simulations facilitate the interpretation of the cation strains determined from the experimental observations. Atomic level alternation of tensile and compressive in-surface strain motifs is caused by the presence of oxygen vacancies on the surface. These strain motifs appear on all surfaces regardless of the processing condition showing that oxygen vacancies are always present. Some of these static strain motifs are not associated with fluxional strain. For example, on the initial surface (*Condition A*), oxygen vacancies are in the middle of flat or buckled (111) terraces where the fluxional strain is very small. These can be categorized as stable vacancies since they are part of a relatively stable surface structure.

Under intense electron irradiation (*Condition C*), the $CeO_2$ nanoparticle will undergo reduction and the concentration of oxygen vacancies will increase at surface and bulk sites. Notice that this does not show up as strong strain in the static strain maps because the surface bond length is normalized to the bulk which will also be reduced. Essentially the vacancy concentration is higher but also more homogeneous in the reduced system minimizing the difference between the surface and bulk strains. Also, with a high vacancy concentration, the tensile and compressive strain fields may partially cancel, reducing the atomic level strain heterogeneity as observed in the strain map. Both these effects explain why the strain magnitudes are mostly lower in **figure 4** compared to **figure 2** and **3**. However, the higher concentration of vacancies in *Condition C* also causes destabilization of the surface cation lattice leading to surface transport (**see supplemental info; cation column migration, figure S5**). The terrace marked "missing columns" in **figure 4a** has shortened due to cations detaching from the step edge. The current step edge associated with the missing columns shows very high static and fluxional strain suggesting that another detachment event may be imminent. This may indicate that in reducible oxide surfaces, high degrees of static and fluxional strain may be precursors to cation migration.

For all conditions, the step edge sites always show a high degree of fluxionality. These represent unstable anion sites with oxygen vacancies constantly being created and annihilated. The activation energy for creating vacancies at these sites is 0.6 eV or lower and so they will be active even at room temperature (5 – 20 creation/annihilation events per second) [18].

The image recorded in an $O_2$ atmosphere (*Condition B*) shows very diffuse surface contrast almost everywhere and the step edges are barely visible. This highly diffuse contrast arises from very high fluxional strain. The enhanced vacancy creation and annihilation is associated with the presence of molecular oxygen in the environment. Molecular oxygen does not typically dissociate at room temperature [49] and it is not clear if molecular splitting is occurring although it is possible that the electron beam may facilitate molecular dissociation. What is clear is that the presence of oxygen at the surface has a destabilizing influence on the surface structure resulting in the high degree of fluxionality. Further investigation is required to fully understand the surface chemistry taking place when oxygen is present. Surface space charge layers can also produce surface strain fields due to the presence of defects and can alter the strain patterns depending upon the defect concentration [50]–[53] but the relative magnitude of such effects have not been considered in the analysis presented here.

Collectively, the observations show that the $CeO_2$ surface is characterized with a highly inhomogeneous cation sublattice strain fields under both oxidizing and reducing condition. Alternating tensile and compressive strain giving bond length changes of up to +/- 10% are present between adjacent surface cations. Both stable and unstable oxygen vacancy sites co-existing under all conditions suggesting that such highly heterogeneous strain fields will be present at the surfaces for most applications.

## Conclusion:

We have used aberration corrected environmental transmission electron microscopy to visualize, at the atomic level, the surface and subsurface strain fields present in a reducible oxide (ceria) under different redox conditions. Static strain mapping was carried out on the cation sublattice, and this strain was interpreted in terms of the distribution and dynamics of anion vacancies. DFT based simulations of single and double vacancies at the surface along with the TEM image simulations were employed to help elucidate the relationship between anion vacancies and cation sublattice strain. The dynamics or fluxionality of cation columns is captured by the fluxional strain and is associated with the oxygen vacancy creation/annihilation process. High degrees of fluxional strain were visible at the most active surface sites (step-edge sites) where lattice oxygen exchange occurs. Fluxionality is indicative of the presence of unstable oxygen vacancies at those sites. Alternating tensile/compressive static strain fields are associated with stable oxygen vacancies which were present at the surface and subsurfaces of the nanoparticle under all conditions, their density changed depending upon the redox conditions.

## Acknowledgement:

We gratefully acknowledge support from NSF DMR-1308085 (ELL and PAC), DMR-1840841 (TB and PAC), and DOE Basic Energy Science DE-SC0004954 (PH and PAC) and Arizona State University's John M. Cowley Center for High Resolution Electron Microscopy for microscope use. We also thank Gatan, Inc. and B. Miller for the loan and installation of a K2 IS direct electron detection camera.

## Data availability:

The data supporting the findings of this study are available within the article and its Supplementary Information file. Other data are available from the authors upon request.